\documentclass[%
 reprint,
superscriptaddress,
preprintnumbers,
nofootinbib,
 amsmath,amssymb,
 aps,
floatfix,
]{revtex4-1}

\usepackage[colorlinks]{hyperref}
\usepackage{graphicx}
\usepackage{dcolumn}
\usepackage{bm}
\usepackage{color}
\usepackage[dvipsnames]{xcolor}
\usepackage{placeins}
\usepackage{slashed}
\usepackage{multirow}
\usepackage{tabularx}
\usepackage{mathtools}
\usepackage{float}
\usepackage{blindtext}
\usepackage{soul}
\usepackage{slashed}
\usepackage{braket}

\usepackage{amssymb}
\usepackage{amsmath}
\usepackage{bbold}

\usepackage[export]{adjustbox}

\usepackage[caption=false,labelfont=normalsize,labelformat=empty,textfont=normalsize,justification=centering]{subfig}

\hypersetup{
  colorlinks=true,
  linkcolor=red,
  citecolor=blue,
  urlcolor=blue,
  hyperfootnotes=true
}

\definecolor{coral}{RGB}{255,127,80}
\definecolor{indigo}{RGB}{75,0,130}
\definecolor{red}{rgb}{0.9, 0,0}
\definecolor{cerulean}{rgb}{0., 0.62,0.9}
\definecolor{navy}{rgb}{0.05, 0.05,0.8}

\newcommand{\GeV}{{\rm \, GeV}}
\newcommand{\MeV}{{\rm \, MeV}}
\newcommand{\keV}{{\rm \, keV}}

\begin{document}

\preprint{TTP24-009, P3H-24-026}
\title{Axion Dark Matter  from Heavy Quarks}

\author{Mohammad Aghaie}
\affiliation{Dipartimento di Fisica E. Fermi, Universit\`a di Pisa and INFN-Pisa, Largo B. Pontecorvo 3, I-56127 Pisa, Italy}

\author{Giovanni Armando}
\affiliation{Dipartimento di Fisica E. Fermi, Universit\`a di Pisa and INFN-Pisa, Largo B. Pontecorvo 3, I-56127 Pisa, Italy}

\author{Angela Conaci}
\affiliation{Dipartimento di Fisica E. Fermi, Universit\`a di Pisa and INFN-Pisa, Largo B. Pontecorvo 3, I-56127 Pisa, Italy}
\affiliation{Dipartimento di Fisica, Universit\`a della Calabria and INFN-Cosenza, Arcavacata di Rende, I-87036 Cosenza, Italy}

\author{Alessandro Dondarini}
\affiliation{Dipartimento di Fisica E. Fermi, Universit\`a di Pisa and INFN-Pisa, Largo B. Pontecorvo 3, I-56127 Pisa, Italy}
\affiliation{Institute for Theoretical Particle Physics, KIT, 76128 Karlsruhe, Germany}

\author{Peter Mat\'ak}
\affiliation{Department of Theoretical Physics, Comenius University in Bratislava, Mlynsk\'a dolina, 84248 Bratislava, Slovak Republic}

\author{Paolo Panci}
\affiliation{Dipartimento di Fisica E. Fermi, Universit\`a di Pisa and INFN-Pisa, Largo B. Pontecorvo 3, I-56127 Pisa, Italy}

\author{Zuzana \v{S}insk\'{a}}
\affiliation{Department of Theoretical Physics, Comenius University in Bratislava, Mlynsk\'a dolina, 84248 Bratislava, Slovak Republic}

\author{Robert Ziegler}
\affiliation{Institute for Theoretical Particle Physics, KIT, 76128 Karlsruhe, Germany}
\affiliation{Institute for Theoretical Physics, Heidelberg University, 69120 Heidelberg, Germany}

\begin{abstract}
We  propose simple scenarios where the observed dark matter abundance arises from decays and scatterings of heavy quarks through freeze-in of an axion-like particle with mass in the $10 \keV - 1 \MeV$ range. These models can be tested by future X-ray telescopes, and in some cases will be almost entirely probed by searches for two-body decays $K \to \pi + {\rm invis.}$ at NA62. As a byproduct, we discuss the cancellation of IR divergencies in flavor-violating scattering processes relevant for thermal axion production, and derive the general contribution to axion-photon couplings from all three light quarks. 
\end{abstract}

\maketitle

\section{Introduction}

 QCD Axions and axion-like particles (ALPs) with masses below the MeV scale  are excellent Dark Matter (DM) candidates, provided that the associated Peccei-Quinn (PQ) breaking scale $f_a$ is sufficiently large in order to ensure stability on cosmological scales. These particles are light enough to be produced in stellar plasmas, and constraints from star cooling typically require $f_a \gtrsim 10^9 \GeV$, seemingly rendering axion production at particle colliders hopeless. However, collider searches are actually sensitive to such large scales, if the axion has flavor-violating (FV) couplings to SM fermions~\cite{Feng:1997tn, Kamenik:2011vy, Bjorkeroth:2018dzu, MartinCamalich:2020dfe, Calibbi:2020jvd, Ziegler:2023aoe}. Precision flavor experiments then allow to probe scales of the order of  $ f_a \sim 10^{12}$ GeV by searching for  $K \to \pi + {\rm invis.}$ at NA62~\cite{Goudzovski:2022vbt}, $10^{10} \GeV$ with $\mu \to e + {\rm invis.}$  at MEG-II~\cite{Calibbi:2020jvd, Jho:2022snj}, Mu3e~\cite{Knapen:2023zgi}, Mu2e or COMET~\cite{Hill:2023dym}, and $10^{9}
\GeV$ with $B \to K + {\rm invis.}$  at Belle II~\cite{MartinCamalich:2020dfe}. 

As flavor-violating axion couplings are determined by the misalignment of PQ charges and SM Yukawas, their prediction from UV scenarios requires a theory of flavor. Particularly  economic models of this kind can be constructed when PQ acts as a flavor symmetry explaining Yukawa hierarchies~\cite{Davidson:1981zd, Wilczek, Berezhiani:1989fp, Flaxion, Calibbi:2016hwq}, although the resulting size of flavor-violating couplings largely depends on the particular scenario. Here instead we link the size of flavor-violating axion couplings to the observed DM relic abundance, requiring
thermal production of DM axions in the right amount via freeze-in of decays (and scatterings) of SM fermions. As $\Omega_a \propto m_a \Gamma (f_i \to f_j a)$, this fixes the rate of these decays for a given axion mass, which indeed is in the reach of current experiments.  This idea has originally been proposed in Ref.~\cite{Panci:2022wlc} in the context of lepton flavor-violating (LFV) decays, it is the purpose of this article to extend the analysis to quarks.

The quark scenario differs from the lepton scenarios in several aspects. Thermal axion production has to respect the Warm DM bound, $m_a \gtrsim 10 \keV$, and axion decays into photons have to be sufficiently suppressed in order to satisfy stringent constraints from X-ray and low-energy $\gamma$-ray line searches. This requires the absence of EM and color anomalies, so that the decay rate is additionally suppressed by power of $m_a^4/m_f^4$. In the case of LFV decays $f = e, \mu$, so that some hierarchy between diagonal and off-diagonal couplings  is needed in order to ensure sufficient stability. In contrast  
 in the quark case the mass suppression is at least $m_a^4/m_\pi^4$, thus improving axion stability and reducing the need of coupling hierarchies. Another important difference is the relative size of axion production rates from decays and scattering processes.
 While in the LFV scenarios production from diagonal scattering is suppressed with respect to decays by a factor of $\alpha_{\rm em}$, in the quark case this becomes a factor $\alpha_s$, so that decays and scattering are almost equally relevant for couplings of similar size. This also implies that NLO corrections are sizable, and we will discuss the corrections from flavor-violating scattering processes, which naively involves IR divergences. However, we will demonstrate that such terms are cancelled in the relevant temperature regime by thermal and virtual corrections, partially reproducing results in Ref.~\cite{Czarnecki:2011mr}.   
 
We now proceed by defining the basic framework in Section~\ref{framework} and introduce two classes of simple benchmark scenarios. We then discuss axion stability, axion production in the early universe and constraints  from structure formation and astrophysics. We use these results to project present constraints and future sensitivities on the 2-dimensional parameter of our benchmark models in Section~\ref{results}, before concluding in Section~\ref{conclusions}.
\section{Framework}
\label{framework}
We consider an anomaly-free ALP $a$ that only couples  to SM quarks 
\begin{align}
    \mathcal{L}=\frac{1}{2}(\partial_{\mu}a)^2-\frac{m_a^2}{2}a^2+\frac{\partial_{\mu}a}{2f_a}\overline{q}_i\gamma^{\mu} \left(C^V_{q_i,q_j}+C^A_{q_i,q_j}\gamma_5 \right)q_j,
    \label{lag}
\end{align}
where $C^{V,A}_{q_i,q_j}$ are traceless hermitian matrices in flavor space. These couplings arise  from the misalignment of PQ charges and quark Yukawa matrices 
\begin{align}
    \begin{split}
        &C^{V,A}_q=U^{\dagger}_{q_R}X_{q_R}U_{q_R}\pm U^{\dagger}_{q_L}X_{Q_L}U_{q_L}\, ,
        \end{split}
\end{align}
where $q=u,d$, and $X_{Q_L}, X_{u_R}, X_{d_R}$ are traceless diagonal matrices containing the PQ charges of $Q_L, u_R, d_R$, respectively, while $U_{q_L, q_R}$ are unitary matrices that diagonalize the quark Yukawas according to $Y_q^{\text{diag}}=U_{q_L}^\dagger Y_q U_{q_R}$, where $V_{\rm CKM} = U_{u_L}^\dagger U_{d_L}$ is the CKM matrix.

\smallskip
Different scenarios can occur depending on the specific choices for  flavor rotations $U_q$ and PQ charges $X_q$. In the following we consider two classes of  benchmark scenarios. In the first class we take into account a single flavor transition at a time, so only two charges of  right-handed (RH) quarks are different from zero, e.g. $X_{d_R} = {\rm diag} (0,1,-1), X_{u_R} = X_{Q_L} =0$. The corresponding unitary matrix is restricted to a rotation in the same sector, i.e. is a rotation in the 2-3 plane by some angle $\alpha$ with $0 \le \alpha \le \pi/2$. This gives 
  \begin{align}
  \label{models}
        C_d^{V} & =C_d^{A} = \begin{pmatrix}
    0 &0 & 0 \\
0 & \sin{\alpha} & \cos{\alpha}\\
0 & \cos{\alpha} & -\sin{\alpha}  
\end{pmatrix} \, , & C^{V}_{u}  & = C^A_u  = 0 \, .
    \end{align}
We call this scenario the ``$bs$ scenario", analogously we define the $bd$, $cu$, $sd$, $tu$ and $tc$ scenarios.  These benchmarks  scenarios have only three free parameters: the ALP mass $m_a$, the decay constant $f_a$, and the rotation angle $\alpha$ that controls the ratio of flavor-diagonal and off-diagonal ALP couplings. We will fix one of these parameters ($f_a$) by demanding that  ALPs are thermally produced in the right abundance via thermal freeze-in. As a consequence, we will obtain a two-dimensional parameter space in the plane ($m_a, \alpha$), which is subject to various constraints from direct searches, astrophysics and cosmology. As we are going to discuss in Section~\ref{results}, only few of the six possible scenarios are viable and give rise to a distinct phenomenology. 

The second class of scenarios is obtained by assuming that the unitary flavor rotations are given by the CKM matrix, while PQ charges in the quark sector are either vanishing or taken to be the most general assignment, $X_q = {\rm diag} (1, X, -1-X)$, where the PQ charge $X$ is a real number. Below we consider two explicit benchmark scenarios: either only left-handed quarks are charged under PQ, i.e., $X_{u_R} = X_{d_R} = 0, X_{Q_L} = {\rm diag} (1, X, -1-X)$, and 
the CKM is coming entirely from the down-quark sector,
$U_{u_L} = 1, U_{d_L} = V_{\rm CKM}$, or only right-handed down quarks are charged, $X_{Q_L} = X_{u_R} = 0, X_{d_R} = {\rm diag} (1, X, -1-X)$ and the relevant rotation is CKM-like, $U_{d_R} = V_{\rm CKM}$.
We call these scenario the ``CKM$_{Q_L}$ scenario" and the ``CKM$_{d_R}$ scenario", respectively. These scenarios are considered to be representative for the phenomenology of  more realistic models, where flavor-rotations are determined by the same dynamics that explain fermion mass hierarchies, which may be the PQ symmetry itself~\cite{Flaxion, Calibbi:2016hwq, Linster:2018avp}. As in the first class, these two scenarios have just three parameters, where again $f_a$ is determined by the observed relic abundance, leaving a two-dimensional parameter space in the plane ($m_a, X$), which is subject to phenomenological constraints. 

\subsection{Dark Matter Stability}\label{stab}
To be stable on cosmological scales, axion decays into SM particles must be sufficiently suppressed.  We will take $m_a \ll m_\pi$,  so that only decays into photons are possible. The dominant constraints on the decay rate come from X-ray telescopes, and are of the order $\tau_{\gamma\gamma}\gtrsim (10^{26} \div 10^{28})\, \text{sec}$, depending on $m_a$, which is roughly 10 orders of magnitude larger than the age of the Universe. 

In our benchmark models, the decay $a\to\gamma\gamma$ takes place through quarks loops. For heavy quarks one can use perturbative results, while for lights quarks ($u,d,s$) one has to rely on chiral perturbation theory since $m_a \ll \Lambda_{\rm QCD}$.  In the following we use and extend the results of Ref.~\cite{Bauer:2017ris}. The decay rate into photons is given by
\begin{align}
        \Gamma_{\gamma\gamma}=\frac{\alpha_{\rm em}^2 m_a^3}{64 \pi^3 f_a^2} \left|C^{\rm heavy}_{\gamma\gamma} + C^{\rm light}_{\gamma\gamma}\right|^2
        \label{rate} \, , 
    \end{align}
    where the effective photon couplings receive contributions from heavy and light quarks. The heavy quark contribution is given by 
\begin{align}
\label{heavy}
C^{\rm heavy}_{\gamma\gamma} \approx \sum_{i = c,b,t}  Q_i^2 C_{i}  \frac{m_a^2}{4 m_i^2} \, , 
\end{align}
    where $C_i \equiv C_{q_i q_i}^A$ and we have neglected terms of order $m_a^4/m_i^4$. Using the results detailed in Appendix~\ref{appendix}, the light quarks contribute dominantly through axion-$\pi$, axion-$\eta$ and axion-$\eta^\prime$ mixing
\begin{align}
\label{light}
C^{\rm light}_{\gamma\gamma} & \approx \frac{C_u-C_d}{2}\frac{m_a^2}{m_{\pi}^2}+\frac{\sqrt{2}}{6}(C_u+C_d-C_s)\frac{m_a^2}{m_{\eta}^2} \nonumber \\
     &+\frac{\sqrt{2}}{3}(C_u+C_d+2 C_s)\frac{m_a^2}{m_{\eta'}^2}  \, , 
\end{align}
with $(m_\pi, m_\eta, m_{\eta^\prime}) = (135, 548, 958) \MeV$,  
and we have neglected multiplicative corrections of order $m_a^2/m_{\pi,\eta,\eta^\prime}^2$, besides small corrections from isospin breaking.

Thus the effective coupling to photons is suppressed by at least $m_a^2/m_\pi^2$, since there is no color nor electromagnetic anomaly~\cite{Nakayama:2014cza, Takahashi:2020bpq, Han:2020dwo,  Han:2022iig, Sakurai:2022roq}.  As a result the axion lifetime is given by (assuming that axion-pion mixing is the dominant contribution)
\begin{align}
\label{lifetime}
\tau_a \approx 3 \times 10^{26} {\rm sec} \left( \frac{0.1 \MeV}{m_a} \right)^7 \left( \frac{f_a/(C_u - C_d)}{10^9 \GeV} \right)^2 \, , 
\end{align}
so that for parameters consistent with freeze-in production and WDM bounds (see below) the axion lifetime easily exceeds the age of the universe, and can be sufficiently large in order to satisfy the stringent limits from X-ray telescopes. 

 For axion masses in the keV-MeV range we use the constraints summarized in Appendix~A of Ref.~\cite{Panci:2022wlc}, where the strongest bounds are set by different X-rays and low energy gamma rays line searches: Chandra~\cite{Watson:2011dw, Horiuchi:2013noa}, Newton-XMM~\cite{Foster:2022ajl}, NuStar \cite{Perez:2016tcq,Roach:2019ctw,Ng:2019gch,Roach:2022lgo}, and INTEGRAL~\cite{Laha:2020ivk}. For heavier masses $1\, \text{MeV} \lesssim m_a  \lesssim 1000$ MeV the most stringent limits on the $a\to\gamma\gamma$ decay rate come from COMPTEL and EGRET, and we take the constraints presented in Ref.~\cite{Essig:2013goa}. Further limits are provided by the optical depth since recombination, which is measured by the Planck collaboration~\cite{Planck:2018vyg}. Fast DM decay into photons would significantly modify the fraction of free electrons after reionization and, consequently, would attenuate the small-scale acoustic peaks of the CMB power spectrum. The model-independent bounds for the optical depth can be found in Ref.~\cite{Cirelli:2009bb, Liu:2016cnk}. These limits constrain rates of the order $\tau_{\gamma \gamma}\approx 10^{24}$ sec, and are therefore less constraining than the X-ray telescopes in the relevant parameter region (cf. Fig.~\ref{res}). Various future X-ray missions  are designed to further extend the limits,  and we use the optimistic projections collected in Ref.~\cite{Panci:2022wlc} for GECCO~\cite{Coogan:2021rez}, THESEUS~\cite{Thorpe-Morgan:2020rwc} and Athena~\cite{Neronov:2015kca,Dekker:2021bos,Ando:2021fhj}.

\label{sec:IRfiniteness}
\begin{figure*}[tbp]
\subfloat{\label{fig1a}}
\subfloat{\label{fig1b}}
\subfloat{\label{fig1c}}
\subfloat{\label{fig1d}}
\subfloat{\label{fig1e}}
\subfloat{\label{fig1f}}
\centering\includegraphics[scale=1]{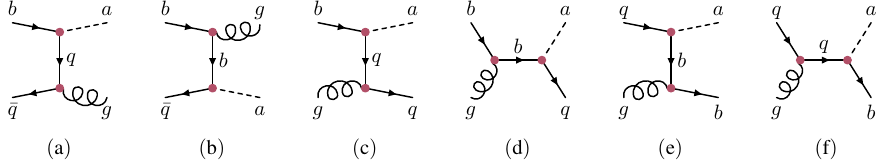}
\caption{\label{fig1} Diagrams contributing to axion production from flavor-violating quark and gluon scatterings.}
\end{figure*}

\subsection{Dark Matter Production}\label{prod}
For sufficiently large decay constants, $f_a \gtrsim 10^8 \GeV$, the axion was never in thermal equilibrium with the SM bath. Thermal axions are then produced via $2 \to 2$ scattering and decay processes of quarks in the thermal bath, which allows to explain the observed DM relic density through the {\it freeze-in} mechanism~\cite{Hall:2009bx}. The total relic axion abundance is given by $\Omega_a h^2= \Omega_a h^2|_{\rm dec}+\Omega_a h^2|_{\rm scatt}$, where $\Omega_a h^2|_{\rm dec}$ and  $\Omega_a h^2|_{\rm scatt}$ are the contributions from (flavor-violating) quark decays $q_i\to q_j a$ and  flavor-diagonal quark scattering processes $q_i g (\gamma) \to q_i a$ and  $q_i \overline{q}_i \to g (\gamma)\,a$,   respectively (we will comment on flavor-violating scattering processes in Section~\ref{sec:IRfiniteness}). The corresponding cross-sections read 
\begin{eqnarray}
\sigma_{q_i \gamma \to q_i a} & = & \frac{\alpha_{\rm em} Q_i^2}{8 f_a^2} |C^A_{q_i q_i}|^2 \frac{x \left(-2 \ln x - 3 + 4 x - x^2\right)}{1-x} \, , \nonumber \\
\sigma_{q_i \overline{q}_i \to \gamma a} & = & \frac{\alpha_{\rm em} Q_i^2}{f_a^2} |C^A_{q_i q_i}|^2 \frac{x \tanh^{-1} (\sqrt{1-4x})}{1-4x} \, , 
\end{eqnarray} 
where $x= m_{q_i}^2/s$ and $Q_i$ is the electric charge of $q_i$. The corresponding gluon scattering processes are obtained from these results by replacing $\alpha_{\rm em} Q_i^2 \to \alpha_s/6$ in $\sigma_{q_i \gamma \to q_i a} $ and  $\alpha_{\rm em} Q_i^2 \to 4 \alpha_s/9$ in $\sigma_{q_i \overline{q}_i \to \gamma a}$, in agreement with e.g. Ref.~\cite{Arias-Aragon:2020shv}. The decay rate is given by 
\begin{eqnarray}
\label{FVdecay}
\Gamma_{q_i \to q_j a}  & = & \frac{m_{q_i}^3}{64 \pi f_a^2} \left|C_{q_i q_j}\right|^2  \left( 1- \frac{m_{q_j}^2}{m_{q_i}^2}\right)^3 \, ,
\end{eqnarray}
where $C_{q_i q_j}\equiv\sqrt{|C^V_{q_i q_j}|^2+|C^A_{q_i q_j}|^2}$ and we have neglected the ALP mass. Following Refs.~\cite{Hall:2009bx, Belfatto:2021ats}, one can use these results to derive analytical estimates for the corresponding contributions to the {\it freeze-in} abundance, assuming that the effective number of relativistic degrees of freedom in the SM bath is approximately constant and that the axion production takes place during radiation domination. Under these assumptions one obtains, including charge multiplicities (cf,  Appendix C in Ref.~\cite{Badziak:2024szg})
\begin{widetext}
\begin{eqnarray}
\label{abundancedecays}
     \Omega_a h^2|_{\rm dec} & \approx & 0.12\left(\frac{mx_a}{0.1 \MeV}\right)\left(\frac{9.7\times 10^{9}\text{GeV}}{f_a/C_{q_iq_j}}\right)^2 \left(\frac{m_{q_i}}{\text{GeV}}\right)\left(\frac{70}{g_*(m_{q_i})}\right)^{3/2} \qquad \mbox{for decays}\, , \\
     \Omega_a h^2|_{\rm scatt} & \approx & 0.12 \left(\frac{m_a}{0.1 \MeV}\right)\left(\frac{1.4\times10^{10}\text{GeV}}{f_a/C^A_{q_i q_i}}\right)^2 \left(\frac{m_{q_i}}{\text{GeV}}\right)\left(\frac{70}{g_*(m_{q_i})}\right)^{3/2} \left( \frac{\alpha_s (m_{q_i})}{0.48} \right) \qquad \mbox{for  scattering} \, ,
     \label{abundancescatt}
    \end{eqnarray}
    \end{widetext}
where we have omitted the sub-dominant contribution from photon scattering.

It is clear from Eqs.~\eqref{abundancedecays} and \eqref{abundancescatt} that the scattering contribution is only slightly smaller than the contribution from quark decays, as a result of the large size of the strong coupling close to the GeV scale. This also implies that omitting higher-order QCD corrections is not a good approximation, so we consider our leading-order results to be valid only up to ${\cal O}(1)$ corrections, which however only has a mild impact on the relevant model parameter $f_a$. Keeping in mind this uncertainty, we can still obtain more accurate expressions by solving the Boltzmann equation numerically, which leads to the results presented in Section \ref{results}.  We stress that for this procedure we only use the temperature dependence of energy and entropy degrees of freedom $g_{(s)*}(T)$, and unlike Ref.~\cite{DEramo:2021usm} we neither consider thermal masses nor flavor off-diagonal scattering processes. Indeed both effects represent only a subset of the full (and unknown) NLO corrections to the leading order effects to which we restrict here, as explained in more detail in Section~\ref{sec:IRfiniteness}. 

In addition to the purely IR contribution to the DM abundance discussed above, {\it freeze-in} scenarios are potentially sensitive also to processes that are dominated by high temperatures. In particular, at energies above the electroweak scale, we have to take into account also operators like
\begin{align}
\label{Higgs}
    \mathcal{L}_{\rm eff}=- C^A_{q_iq_j} \frac{i a}{f_a}\frac{m_{q_i}}{v}H \overline{Q}_i q_{Rj} \, , 
\end{align}
where $Q_i$ ($H$) denotes the quark (Higgs) doublet field. This operator  can be obtained from Eq.~\eqref{lag} upon integrating by parts and using the equations of motion in the unbroken phase, and for simplicity we have set $C^V_{q_iq_j} = C^A_{q_iq_j}$. This gives rise to  scattering processes like $ \overline{q}_i q_j \to h a$, which lead to axion production rates that are UV sensitive and thus depend on the reheating temperature $T_R$. The corresponding UV contribution to the relic abundance can be related to the decay contribution as~\cite{Hall:2009bx}
\begin{align}
    \Omega_a h^2|_{\rm UV} \approx \frac{m_{q_i}T_R}{3\pi^3 v^2} \times \Omega_a h^2|_{q_i\to q_j a} \, .
\end{align}
While one could take into account such UV sensitive contributions on the price of introducing $T_R$ as an additional parameter of the models, here we want to stick to the minimal number of parameters and thus take $T_R$ sufficiently small such that the IR contributions always dominate the relic abundance. As we will see below, this procedure also suppresses the misalignment contribution, which also depends on additional parameters (the original misalignment angle). Hence, we establish an upper bound on $T_R$ by requiring that the axion abundance generated from UV sensitive processes is smaller than the one from decays, giving $T_R<3\pi^3v^2/m_{q_i} = 3 \times 10^6 \GeV (\GeV/m_{q_i})$. Equivalently we can consider an upper bound on the Hubble parameter at reheating
\begin{align}
    H_R < 11 \keV \left(\frac{\GeV}{m_{q_i}}\right)^2 \, , 
\end{align}
where have assumed $g_*(T_R)\approx 106.75$.


We can now discuss possible sources of non-thermal production. The most relevant is the misalignment mechanism for ALPs~\cite{Arias:2012az, Blinov:2019rhb}. Also this contribution
depends on the reheating temperature, as  the onset of axion oscillations (defined by\footnote{For the numerical values below we have used $m_a = 1.6 H(Tosc)$ as suggested in Ref.~\cite{Blinov:2019rhb}.} $m_a \simeq H$) occurs prior to reheating in the axion mass range under consideration. Todays misalignment abundance is then suppressed due to the dilution that occurred during an initial period of matter domination\footnote{If inflation ends in a period of kination instead the misalignment contribution would be enhanced.} that took place between the onset of oscillations and $T_R$. The resulting ALP abundance in terms of the misalignment angle $\theta_0$ is then given by \cite{Blinov:2019rhb, Visinelli:2009kt, Arias:2021rer}
\begin{align}
  \Omega_a h^2|_{\rm mis}\approx 4 \times 10^{-3} \left(\frac{H_{R}}{11 \keV }\right)^{1/2}\left(\frac{f_a\theta_0}{ 10^{10}\,\text{GeV}}\right)^2 .
\end{align}
We notice that the misalignment contribution can be somewhat larger than in models where the ALP is only coupled to leptons~\cite{Panci:2022wlc}, as heavy quarks require sizable $f_a$ scales. Still,  misalignment production is never relevant in the interesting parameter region compatible with astrophysical bounds, as we will discuss in Sec.~\ref{results}.

\subsection{Flavor-violating Scattering Processes and Infrared Finiteness}

In addition to $q_i\rightarrow q_j a$ decays, axion production via freeze-in is also affected by flavor-changing quark and gluon scatterings \cite{Arias-Aragon:2020shv, DEramo:2021usm, DEramo:2023asj}. For the special case of $b\bar{q}\rightarrow ag$, $b g\rightarrow aq$, and $qg\rightarrow ab$ reactions, with $q=s,d$, the diagrams are shown in Fig.~\ref{fig1}. The squared amplitude for any of these three processes depends both on the bottom and the light quark masses. Neglecting the latter, an infrared divergence occurs, while using non-zero light-quark mass may still lead to an unphysical enhancement of the cross section due to the sizeable quark-mass hierarchy. Let us consider the square of the diagram in Fig. \ref{fig1a}. If $q$ is massless and its momentum is collinear to the momentum of the gluon, a singularity occurs. In order to isolate the divergent terms, we assign the light quark a small mass $m_q$ as a regulator, and take the limit $m_q \to 0$ when possible.
The square of the diagram summed over spins and integrated over the final state momenta can be expressed as a unitary cut of the forward-scattering diagram
\begin{align}
\label{eq:cut1}
\includegraphics[scale=1,valign=c]{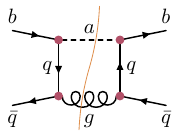}\propto\hskip1mm
2s+\frac{4m^4_b}{s}+\frac{2sm^2_b}{s-m^2_b}\ln\frac{(s-m^2_b)^2}{s m^2_q} \, , 
\end{align}
where $s=(p_{b\vphantom{\bar{q}}}+p_{\bar{q}})^2$ is the squared centre-of-mass energy. A logarithmic divergence occurs in the last term, and the same happens in all contributions involving the diagrams in Figs.~\ref{fig1a} and \ref{fig1c}. 
To deal with these singularities, we follow the procedure introduced in Ref.~\cite{Racker:2018tzw} based on the Kinoshita-Lee-Nauenberg (KLM) theorem \cite{Kinoshita:1962ur, Lee:1964is, Frye:2018xjj}. The forward-scattering diagram in Eq.~\eqref{eq:cut1} allows for two other unitary cuts corresponding to so-called anomalous thresholds \cite{Mandelstam:1960zz, Cutkosky:1961, Goddard:1969ci, Hannesdottir:2022bmo}, which evaluate to
\begin{widetext}
\begin{align}\label{eq:cut2}
\includegraphics[scale=1,valign=c]{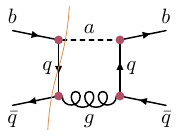}+ \includegraphics[scale=1,valign=c]{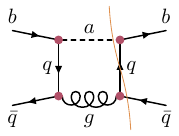}\propto\hskip1mm
-4m^2_b-\frac{4m^4_b}{s}-\frac{2sm^2_b}{s-m^2_b}\ln\frac{(s-m^2_b)^2m^2_b}{s^2m^2_q} \, ,
\end{align}
\end{widetext}
yielding a finite result for $m_q\rightarrow 0$  when added to Eq.~\eqref{eq:cut1}. We apply this procedure whenever an infrared divergence occurs in the total cross sections of the processes listed in Fig.~\ref{fig1}, which allows us to obtain well-defined expressions. Still, another sort of singularity persists. 

The square of the diagram in Fig.~\ref{fig1d} leads to a finite total cross section, but diverges for small gluon energy in thermal averaging.
The problem has been resolved in Ref.~\cite{Czarnecki:2011mr} for electromagnetic corrections to charged particle decays in a thermal medium. In our case, we must include the gluon-induced bottom thermal mass and wave-function renormalization factor in the $b\rightarrow qa$ decay. Furthermore, in analogy to Eq.~(9) of Ref.~\cite{Czarnecki:2011mr}, we add the temperature-dependent part of the gluon Bose-enhancement factor in the $b\rightarrow qag$ decay. As a result, the $s$-channel $bg\rightarrow aq$ cross section is found to be completely canceled for values of $s < 2m^2_b$. Putting everything together, the ratio of temperature-dependent axion production rates from scattering $\gamma_S$ and decays $\gamma_D$ is finite in the limit $m_q \to 0$ and given by
\begin{widetext}
\begin{align}\label{eq:SD}
\frac{\gamma_S}{\gamma_D}=& \frac{\alpha_s}{\pi}\int\displaylimits^\infty_1
d\Tilde{s} \frac{\Tilde{s}-1}{\sqrt{\Tilde{s}}}
\frac{K_1\big(x_b\sqrt{\Tilde{s}}\big)}{K_1\big(x_b\big)}\bigg\{
-\Tilde{s}-4-\frac{3}{\Tilde{s}}+\frac{12}{\Tilde{s}-1}\\
&+\bigg[8-\frac{8}{\Tilde{s}}+\frac{10}{\Tilde{s}-1}+\frac{4}{(\Tilde{s}-1)^2}\bigg]\ln\Tilde{s}
+\bigg[\Tilde{s}+2-\frac{8}{\Tilde{s}-1}+\frac{4}{(\Tilde{s}-1)^2}
\bigg]\theta(\Tilde{s}-2)\bigg\} \, , \nonumber
\end{align}
\end{widetext}
with $\Tilde{s}=s/m^2_b$ and $x_b=m_b/T$. 

Axion production via {\it freeze-in} is dominated by temperatures slightly below the heavy quark mass. From Eq.~\eqref{eq:SD}, taking $x_b=3$ leads to $\gamma_S/\gamma_D\approx 0.7$. This contrasts with the findings of Ref.~\cite{DEramo:2021usm}, where thermal masses are used in the leading order cross-sections to handle divergencies. This leads to an enhancement of the scattering production by two orders of magnitude relative to decay at the relevant temperature slightly below the heavy quark mass. In our procedure, this enhancement is cancelled by the contributions of the anomalous thresholds, as in Eq.~\eqref{eq:cut2}. Those are indeed related to thermal-mass effects, but instead of the scattering, they enter the leading-order $b\rightarrow qa$ decay kinematics through mass-derivative relations (see Eq.~(35) in Ref. \cite{Blazek:2021gmw} for an example).
 However, we note that our approximation of thermal effects is not complete and only represents a minimal set of contributions needed for infrared finiteness. A more complete treatment of thermal corrections may be considered in future work. 

\subsection{Warm Dark Matter}
Soon after its production, DM free-streams and suppresses the primordial fluctuations related to the matter power spectrum. DM free-streaming leaves its footprints on large-scale structures and can be constrained by looking at the absorption features of the spectra of 
distant quasars through the Lyman-$\alpha$ forest (Ly-$\alpha$)~\cite{Boyarsky:2008xj}. In particular, one can set a \lq\lq warmness bound" on the DM mass to avoid large free-streaming \cite{Viel:2013fqw, Baur:2015jsy, Irsic:2017ixq}.  The Ly-$\alpha$ limits have been recasted for different freeze-in processes by computing the exact DM velocity distribution, which results in the ``Warm Dark Matter" (WDM) constraint~ \cite{DEramo:2020gpr, Ballesteros:2020adh, Decant:2021mhj}
\begin{align}
\label{WDMbound}
    m_a\gtrsim 0.01 \MeV \left(\frac{m_{\rm WDM}}{3.5\,\text{keV}}\right)^{4/3}\left(\frac{70}{g^*(m_q)}\right)^{1/3} \, , 
\end{align}
where $m_{\rm WDM}\approx 3.5 \keV$ or $5.3 \keV$ for the conservative and stringent bounds, respectively.

\subsection{Other Astrophysical Bounds}\label{AstroBounds}
Sufficiently light ALPs coupled to SM fermions can efficiently extract energy from stellar objects and are subject to limits from star cooling~\cite{Raffelt:1996wa}. Quark couplings induce axion couplings to nucleons, which allow for efficient axion production in hot stellar plasmas, such as in the proto-neutron star formed during core-collapse supernovae. Sufficiently light axions ($m_a \lesssim 100 \MeV$) would extract energy from the proto-neutron star, which is constrained by the usual energy loss argument for SN1987A~\cite{Raffelt:1996wa}.  Lighter axions ($m_a \lesssim 0.4 \keV$) are also constrained by measurements of the White Dwarf (WD) luminosity function, which primarily limits electron couplings at the order of $f_a/C_e \ge 2.5 \times 10^9 \GeV$~\cite{MillerBertolami:2014rka}, but due to Renormalization Group evolution this also puts constraints on axion couplings to top quarks~\cite{Feng:1997tn} at the level~\cite{MartinCamalich:2020dfe}
\begin{align}
\label{Ct}
f_a/C_t \ge 1.7 \times 10^9 \GeV \, , 
\end{align}
where we ignored the mild logarithmic dependence on the UV scale by setting $f_a = 10^{10} \GeV$ for simplicity. 

In order to extract the resulting SN1987A limits on  axion-quark couplings, we match the Lagrangian in Eq.~\eqref{lag} to the axion-nucleon effective Lagrangian. Following Ref.~\cite{GrillidiCortona:2015jxo}, we obtain the following effective Lagrangian in the non-relativistic limit, which should be reliable as long as the ALP mass and the relevant energies are smaller than the QCD mass gap $\Delta \approx 100$ MeV
    \begin{align}
\mathcal{L}_{aN} & =\overline{N}v^{\mu}\partial_{\mu}N+\frac{\partial_{\mu}a}{f_a}\frac{C_u -C_d}{2} \Delta_{u-d} \overline{N}S^{\mu}\sigma^3 N   \\
 & +\frac{\partial_{\mu}a}{f_a}  \left[ \frac{C_u +C_d}{2} \Delta_{u+d} 
 +  \sum_{q=s,c,b,t} C_q \Delta q  \right] \overline{N}S^{\mu} N \, , \nonumber 
\end{align}
where $N=(p,n)$ is the nucleon isospin doublet, $v^{\mu}$ is the four-velocity of the nucleon, $2S^{\mu}\equiv \gamma^{\mu}\gamma^5$ is the spin operator and $\Delta_{u \pm d} \equiv \Delta u \pm \Delta 
 d$. The coefficients $\Delta q, q=u,d,s,c,b,t$ are extracted from lattice QCD studies and low-energy experiments,  and can be found in Ref.~\cite{GrillidiCortona:2015jxo}. Here we use the recent analysis in Ref.~\cite{Badziak:2023fsc}, giving 
\begin{align}\label{nucleoncoup}
    \begin{split}
        &C_p\approx 0.82 \, C_u - 0.45 \, C_d   -0.052 \, C_s \, , \\
         &C_n\approx 0.82 \, C_d - 0.45 \, C_u   -0.052 \, C_s \, , \\
    \end{split}
\end{align}
where we have neglected the contributions from heavy quarks. Bounds on these couplings can be obtained from the burst duration of the neutrino emission of SN1987A, which yields~\cite{Carenza:2019pxu}
\begin{align}
    0.61 g_{ap}^2+g_{an}^2+0.53 g_{an} g_{ap}<8.26 \times 10^{-19} \, ,
\end{align}
where $g_{ai}\equiv C_i m_i/f_a$. These constraints are roughly comparable to limits that can be derived from  observations of neutron star cooling rates~\cite{Buschmann:2021juv}, and give for $C_p \approx C_n \approx C_N$
\begin{align}
\label{SNbound}
f_a/C_N \gtrsim 1.5 \times 10^9 \GeV \, .
\end{align}
\begin{figure*}[t!]
\begin{center}
    \includegraphics[width=0.47\textwidth]{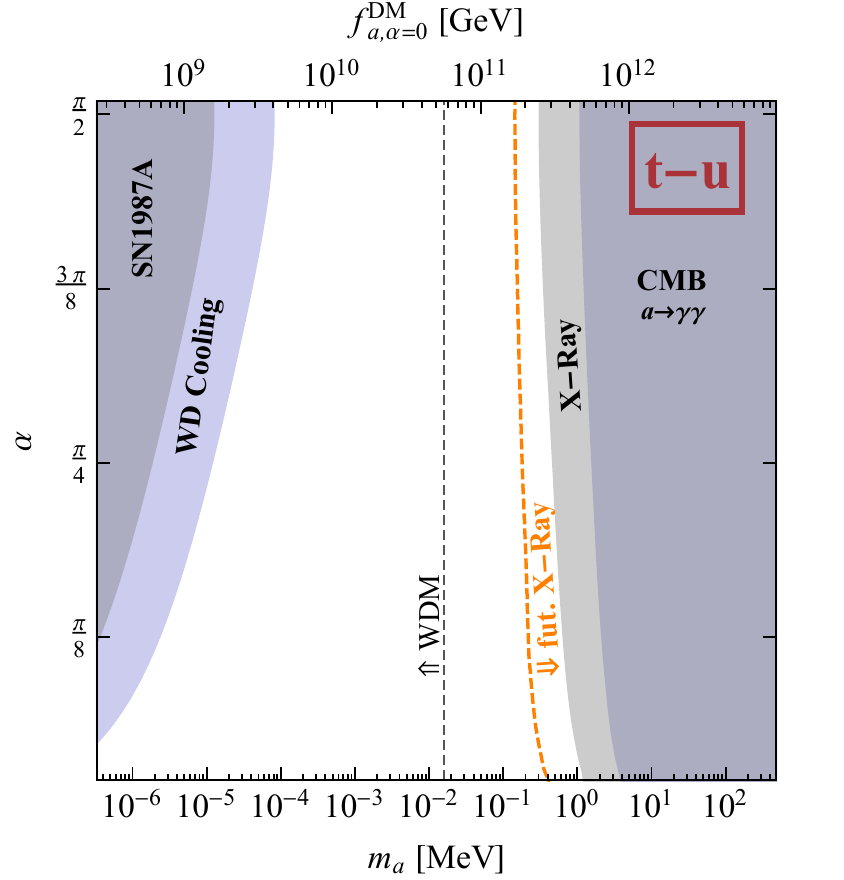}\hfill
	\includegraphics[width=0.47\textwidth]{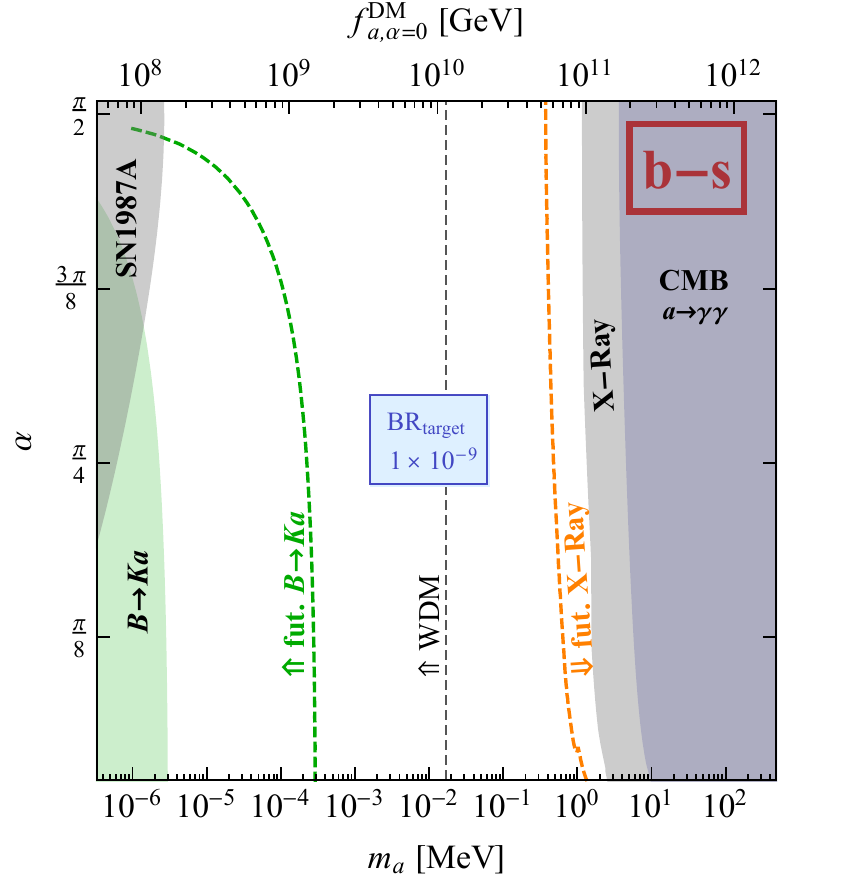}\hfill
   	\caption{Parameter space in the $(m_a, \alpha)$ plane for selected benchmark models defined in Section~\ref{framework}, which reproduce the observed DM relic abundance everywhere. The $tu$ scenario (left panel) is representative for the $tc$ scenario, the $bs$ scenario (right panel) for the $bd$ and $cu$ scenarios. The angle $\alpha$ controls the ratio of flavor-diagonal to flavor-off-diagonal couplings, see Eq.~\eqref{models}. The shaded blue region shows CMB constraints on the axion lifetime, while the shaded gray bound indicates to the region excluded by X-ray telescopes. The dark gray region is excluded by SN (WD) cooling, while the green shaded region is probed by collider searches.  The stringent WDM bound (the conservative is weaker by a factor 1.7) is denoted by a dashed black line  along with the corresponding target branching ratio for $B \to K a$,  and green (orange) dashed lines indicate the prospective limits from future laboratory searches (X-ray telescopes). The upper axis indicates the value of $f_a$ needed to reproduce the relic abundance for $\alpha = 0$, with a mild dependence on $\alpha$.  \label{res}  }
\end{center}
\end{figure*}
\section{Results}\label{results}
It is clear from Eqs.~\eqref{abundancedecays} and \eqref{abundancescatt} that for heavy quarks $m_{q_i} \gtrsim \GeV$ and axion masses $m_a \sim 0.1 \MeV$ satisfying the WDM bound in Eq.~\eqref{WDMbound}, the observed DM relic abundance can be obtained for $f_a \sim 10^{10} \GeV$, while respecting the limits from X-ray telescopes in~Eq.~\eqref{lifetime} and the supernova bounds in~Eq.~\eqref{SNbound}. This also implies that  axion couplings to light quarks, i.e., the $sd$ scenario, are not viable, since the strong constraints from $K^+\to \pi^+ + {\rm inv.}$ searches~\cite{NA62:2021zjw, Goudzovski:2022vbt} essentially exclude the whole parameter space, as the relic abundance require too low $f_a$ values. In contrast  {\it freeze-in} via heavy-quark couplings ($c,b,t$) gives larger values of $f_a$, and there is not much difference between axion production via flavor-diagonal or flavor-violating couplings, as $\alpha_s$ is sizable\footnote{This is in contrast to the $\mu e$ scenario considered in Ref.~\cite{Panci:2022wlc}, where axion production via flavor-diagonal scattering is suppressed by $\alpha_{\rm em}$, and requires very low values of $f_a$  that are already excluded by X-ray searches. These limits disappear in the limit where axion couplings are mainly flavor-violating, which provides a scenario compatible with present laboratory searches for $\mu \to e + {\rm invis.}$ and in the reach of near-future experimental proposals.}. It turns out that all five scenarios in the first class discussed in Section~\ref{framework} are indeed viable for all values of $\alpha$, which controls the ratio of diagonal to off-diagonal couplings. This is because for values of $m_a$ that respect the WDM bound, not only the constraints on flavor-diagonal couplings from star cooling are satisfied, but also the stringent laboratory limits on flavor-violating decays with missing energy are respected, even taking into account near-future projections. This regards $D \to \pi + {\rm inv.}$ searches at CLEO~\cite{CLEO:2008ffk, MartinCamalich:2020dfe}  ($cu$ scenario), $B \to \pi + {\rm inv.}$ searches at BaBar~\cite{BaBar:2004xlo, MartinCamalich:2020dfe} ($bd$ scenario) and $B \to K + {\rm inv.}$ searches at Belle II~\cite{Belle-II:2023esi} and BaBar~\cite{BaBar:2013npw} ($bs$ scenario), where we used the combined limit on the two-body decay recently provided in Ref.~\cite{Altmannshofer:2023hkn}, ${\rm BR} (B \to Ka) < 8.0 \times 10^{-6}$ at 95\% CL for $m_a \ll 100 \MeV$. Note that the $tc$ and $tu$ scenarios are constrained only mildly by WD cooling (Eq.~\eqref{Ct}) and SN1987A (Eq.~\eqref{SNbound}), and SM loop contributions to $K \to \pi a$ are absent in these cases as the axion only couples to RH quarks\footnote{Otherwise one would obtain constraints on $f_a$ of the order of few$\times 10^8 \GeV$~\cite{MartinCamalich:2020dfe}.}. 
\begin{figure*}[t!]
\begin{center}
 \includegraphics[width=0.47\textwidth]{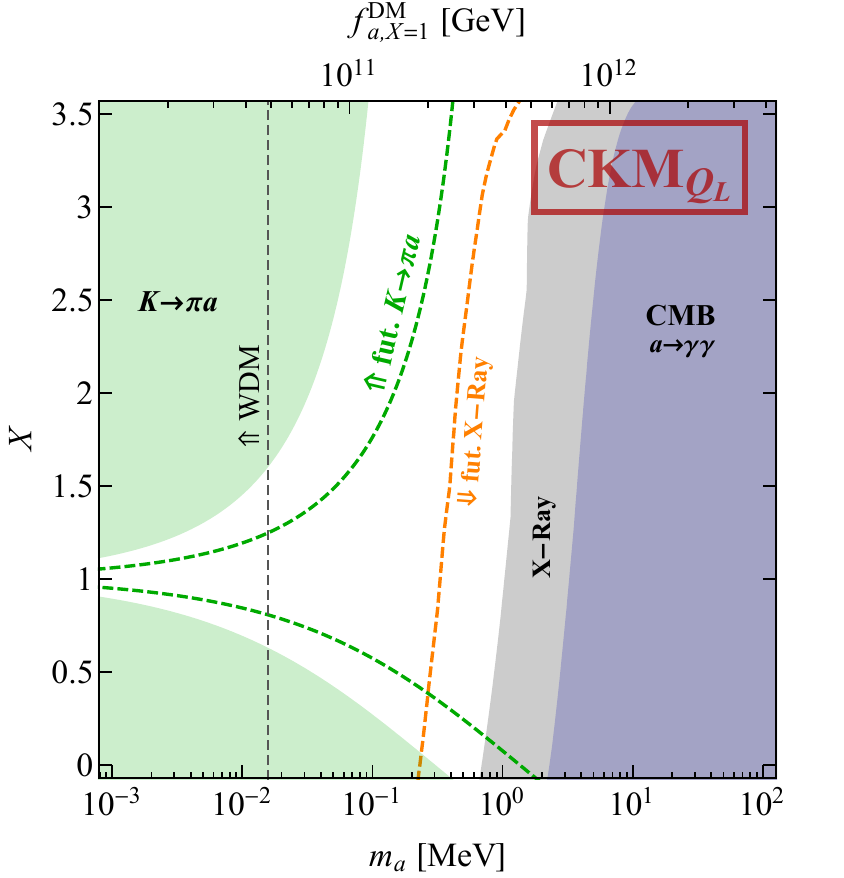}\hfill
    \includegraphics[width=0.47\textwidth]{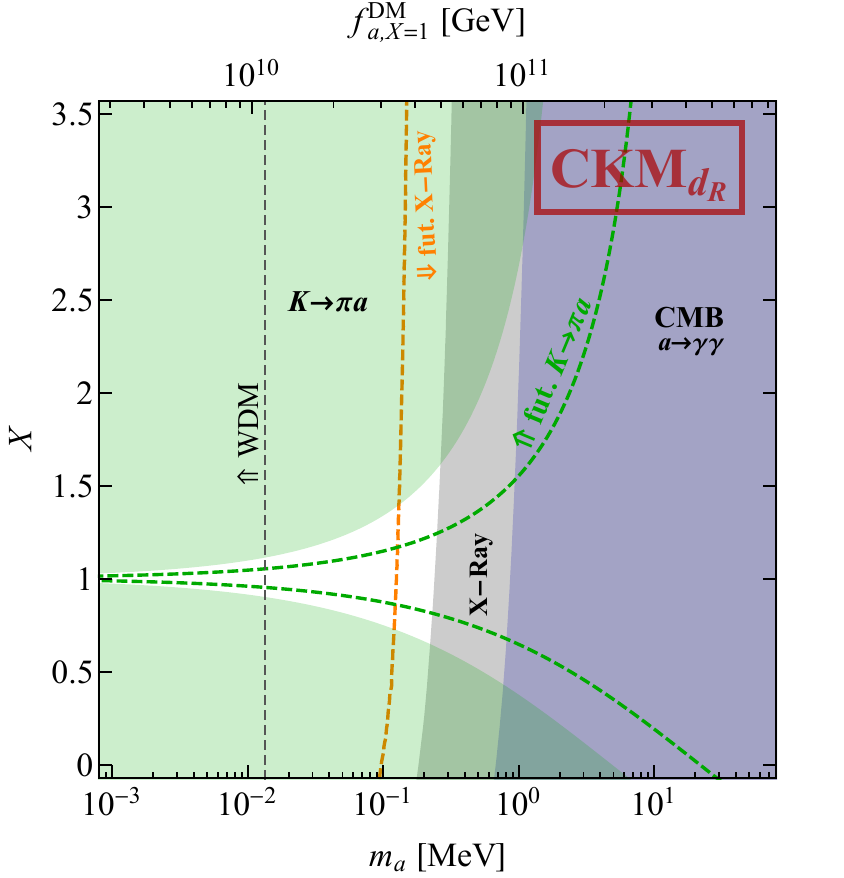}\hfill
	   	\caption{Parameter space in the $(m_a,X)$ plane for the  CKM$_{Q_L}$  (left panel) and  CKM$_{d_R}$ (right panel)  benchmark models defined in Section~\ref{framework}, which reproduce the observed DM relic abundance everywhere. The parameter $X$ controls the ratio of PQ charges for LH quarks (CKM$_{Q_L}$) or RH down-quarks (CKM$_{d_R}$), which are rotated to the quark mass basis by CKM rotations, see Eq.~\eqref{models}. The shaded blue region shows CMB constraints on the axion lifetime, while the shaded gray bound indicates to the region excluded by X-ray telescopes. The green shaded region is excluded by $K \to \pi a$ searches at NA62.  The stringent WDM bound (the conservative is weaker by a factor 1.7) is denoted by a dashed black line  along with the corresponding target branching ratio for $K \to \pi a$,  and green (orange) dashed lines indicate the prospective limits from future laboratory searches (X-ray telescopes). The upper axis indicates the value of $f_a$ needed to reproduce the relic abundance for $X = 1$, with a mild dependence on $X$. \label{Xmodels}}
\end{center}
\end{figure*}
Thus we obtain five simple benchmark models that generate the DM relic abundance and are compatible with all present constraints. In Fig.~\ref{res} we display the 2-dimensional parameter space for two of such models, the $tu$ scenario and the $bs$ scenario. Analogous figures for the $tc$ scenario and the $bd$ and $cu$ scenarios are not shown since they are very similar to the $tu$ and $bs$ scenario, respectively. All scenarios have in common that even future laboratory searches for two-body flavor-violating decays with missing energy will not probe the interesting region of axion masses satisfying the WDM bound, which is right of the vertical dashed black line. For the $bs$ scenario we have indicated the maximal size of the $B \to K a$ target branching ratio compatible with the WDM bound, which is of the order of ${\rm BR} (B \to K a)_{\rm target} \approx 1 \times 10^{-9}$, clearly beyond the reach of running or near-future  $B$-factories~\cite{MartinCamalich:2020dfe} indicated by the green dashed line. The same conclusions are valid for $D \to \pi$, with ${\rm BR} (D \to \pi a)_{\rm target} \approx 2 \times 10^{-10}$, and $ B \to \pi$ transitions with ${\rm BR} (B \to \pi a)_{\rm target} \approx 6 \times 10^{-10}$, such that all scenarios will be tested only by future X-ray telescopes. Note however that colliders are in principle better suited to probe the remaining parameter space as compared to X-ray line searches, as the constrained axion mass scales rather weakly with the axion decay rate into photons, $m_a\propto \Gamma_{a \to \gamma\gamma}^{1/6}$ (Eq.~\eqref{lifetime}), but strongly with flavor-violating decay rates $m_a \propto \Gamma_ {q_i \to q_j a}^{1/2}$ (Eq.~\eqref{FVdecay}), after fixing $f_a^2 \propto m_a$ with the relic abundance (Eq.~\eqref{abundancedecays}).

We finally discuss benchmark models where present collider constraints exceed the WDM bound, such that the remaining parameter space will be complementary probed by  precision flavor experiments and  X-ray telescopes. This is the case for the other two benchmark scenarios discussed in Section~\ref{framework}, where the PQ charge is taken as a free parameter and the rotation to the quark mass basis is fixed by the CKM matrix. 
The resulting parameter space is shown in Fig.~\ref{Xmodels} for the CKM$_{Q_L}$  and the CKM$_{d_R}$ scenario, which reproduce the observed DM relic abundance for the indicated values for $m_a$ and the PQ charge $X$. The dominant contribution to axion production comes from unsuppressed processes involving the heaviest quarks, which is $tt$ scattering in the CKM$_{Q_L}$  scenario and $bb$ scattering in the CKM$_{d_R}$ scenario, since quark mixing in the CKM is small and cannot compensate the mild $\alpha_s$ suppression in scatterings compared to decays. Still the CKM involves a rather large rotation in the $sd$ sector of order $\lambda \approx 0.23$, which induces a sizable coupling of the axion to $sd$ quarks in both scenarios for generic values of $X$, unless the first two generation have the same PQ charge, i.e. $X=1$, leading to an approximate $SU(2)$ symmetry and $C^V_{sd}$ involves additional CKM suppression\footnote{See Ref.~\cite{Linster:2018avp} for a motivated scenario where this situation arises by identifying PQ as a subgroup of a horizontal $U(2)$ symmetry explaining Yukawa hierarchies.}. Close to this value the stringent limits on $K \to \pi a$ from NA62~\cite{NA62:2021zjw} are relaxed, which otherwise give constraints of order $f_a/C^V_{sd} > 4 \times 10^{11} \GeV$~\cite{Goudzovski:2022vbt}. For $|X - 1|  \gtrsim 0.44$ (CKM$_{d_R}$) or $|X - 1|  \gtrsim 0.24$ (CKM$_{Q_L}$) the resulting limits exceed the WDM bound, so that NA62 will probe the remaining parameter space in the near future complementary to future X-ray telescopes. Interestingly, the parameter space of the CKM$_{d_R}$ scenario will be almost entirely probed by $K \to \pi a$ and $a \to \gamma \gamma$ searches, leaving only a narrow region between axion masses $10 \lesssim m_a \lesssim 100 \keV$ and PQ charge $|X - 1| \lesssim 0.15$. 

\section{Conclusions}
\label{conclusions}
To summarize, we have explored the production of axion DM from decays and scatterings of heavy quarks via thermal {\it freeze-in}. This gives rise to very simple scenarios with few parameters able to explain the observed DM abundance, which are subject to various constraints from precision flavor experiments, star cooling, X-ray telescopes and structure formation. Similar to the lepton case explored in Ref.~\cite{Panci:2022wlc}, we have focussed on two classes of models with only two parameters after fixing the axion decay constant to values that reproduce the observed DM relic abundance. Apart from the axion mass, the free parameter is the ratio of flavor-diagonal couplings to flavor-violating couplings in the first class (effective 2-flavor scenarios), and the overall PQ charge in the second, with flavor violation controlled by the CKM matrix.

 Compared to the case of an ALP coupled to electrons~\cite{Panci:2022wlc}, in the quark scenarios the axion decay rate into photons is additionally suppressed by at least a factor $m_e^4/m_{\pi}^4$, which enhances axion stability and eases constraints on flavor-diagonal couplings from X-ray line searches. As axion production from quark scattering is only mildly suppressed with respect to quark decays as a result of large values of $\alpha_s$ close to the GeV scale, we find that there is not much difference between scenarios with flavor-violating coupling and flavor-diagonal couplings of the same size, in stark contrast to LFV models~\cite{Panci:2022wlc}. This also implies that next-to-leading order corrections to axion production are sizable, which we have calculated here for the first time for the case of  flavor-violating $2 \to 2$ scattering processes. We showed that in concordance with the KLN theorem IR divergencies in these processes are cancelled by taking into account  contributions from anomalous thresholds (partially related to thermal corrections), at least for energies below the heaviest quark involved. A more complete analysis of NLO corrections is left for future work.
 
The main results of our analysis are summarized in Fig.~\ref{res} for the effective 2-flavor model and the CKM scenarios in Fig.~\ref{Xmodels}. The allowed parameter space  of all 2-flavor models have similar shapes (thus we only show the $tu$ and $bs$ scenarios as representatives), and are viable except for the $sd$ model, which is essentially ruled out by present $K \to \pi a$ constraints. 
These scenarios will only be probed by future X-ray telescopes, as the sensitivities of future flavor factories will still be  weaker than the constraints on Warm Dark Matter. On the other hand in the CKM scenarios the strongest limits in the low axion mass regime arise from searches for $K \to \pi a$ at NA62. The expected sensitivity together with X-ray line searches will allow to  probe large portions of the remaining parameter space, giving excellent prospects to explore a very simple class of axion DM models at the high-intensity frontier.  

\acknowledgments
We would like to thank Francesco D'Eramo, Kirill Melnikov, Uli Nierste and Diego Redigolo  for useful
discussions. Peter Mat\'{a}k and Zuzana \v{S}insk\'{a} were supported by the Slovak Grant Agency VEGA, project No. 1/0719/23. Peter Mat\'{a}k also received financial support from Slovak Education Ministry contract No. 0466/2022.  The work of Robert Ziegler has received support from the European Union's Horizon 2020 research and innovation programme under the Marie Sk{\l}odowska-Curie grant agreement No 860881-HIDDeN and is partially supported by project B3a and C3b and of the DFG-funded Collaborative Research Center TRR257 ``Particle Physics Phenomenology after the Higgs Discovery''.  The research conducted by Mohammad Aghaie, Giovanni Armando, Alessandro Dondarini, Angela Conaci and Paolo Panci receives partial funding from the European Union–Next generation EU (through Progetti di Ricerca di Interesse Nazionale (PRIN) Grant No. 202289JEW4).
\appendix
\section{Axion-Photon Coupling}
\label{appendix}
In this appendix we present the details of the light quark contribution to the axion-photon coupling in Eq.~\eqref{light} due to axion mixing with $\pi, \eta$ and $\eta^\prime$ in leading order chiral perturbation theory ($\chi$PT). Our analysis complements the results of Ref.~\cite{Ertas:2020xcc}, where the contribution to axion-meson mixing from the axion-gluon coupling was calculated. Here instead we provide the contribution from axion couplings to all three light quarks. 

We start by matching the Lagrangian in Eq.(\ref{lag}) to 3-flavor $\chi$PT. After integrating out the heavy quarks, we define the effective axion couplings as diagonal $3\times3$ matrices $k_{R,L}=1/2 \, {\rm diag} (C^V_{uu} \pm C^A_{uu}, C^V_{dd} \pm C^A_{dd}, C^V_{ss} \pm C^A_{ss})$, and the Lagrangian reads
\begin{align}
    \mathcal{L}_{\text{light}}&=\frac{1}{2}(\partial_{\mu}a)^2-\frac{m_a^2}{2}a^2+\Bar {\Psi}(i\slashed D-M_q)\Psi \\
    &+\frac{\partial_{\mu}a}{f_a}\Bar{\Psi}\gamma^{\mu} \left( k_{L} P_L +k_R P_R \right) \Psi \, , 
\end{align}
where   $\Psi\equiv (u,d,s)^T$ and $M_q = {\rm diag} (m_u, m_d, m_s)$. The chiral Lagrangian is written in terms of the unitary $3\times 3$ matrix $\Sigma$ containing the Goldstone boson octet $(\pi, K, \eta_8)$ and the singlet $\eta_0$ as $\Sigma = \exp{ \left( i \sqrt{2}\Phi/f_\pi \right)}$, where 
\begin{align}
    \Phi =\begin{pmatrix}
        \pi^0 +\frac{\eta_8}{\sqrt{3}} & \sqrt{2} \pi^+ & \sqrt{2} K^+\\
        \sqrt{2}\pi^- & -\pi^0 +\frac{\eta_8}{\sqrt{3}} & \sqrt{2} K^0\\
        \sqrt{2} K^- & \sqrt{2} \bar{K}^0 &-\frac{2}{\sqrt{3}}\eta_8           \end{pmatrix} + \sqrt{\frac{2}{3}}\eta_0 \mathbb{1} \, . 
\end{align}
At leading order the  $SU(3)_{\rm L} \times SU(3)_{\rm R}$ symmetry gives
\begin{align}\label{eq:chirlagAp}
    \mathcal{L}_{\rm \chi PT} & = \frac{1}{2}(\partial_{\mu}a)^2 -\frac{m_a^2}{2}a^2+\frac{f_{\pi}^2}{8}\text{Tr}\left[D_{\mu}\Sigma D^{\mu} \Sigma^{\dagger}\right] \nonumber \\
    & +\frac{f_{\pi}^2}{4}B_0 \text{Tr}\left[M_q\Sigma^{\dagger}+ {\rm h.c.} \right]-\frac{1}{2}M^2_0 \eta_0^2,
    \end{align}
where the explicit mass term $M_0$ takes into account the explicit breaking of the anomalous $U(1)_A$ symmetry and the covariant derivative reads
\begin{align}
     &D_{\mu}\Sigma =\partial_{\mu}\Sigma+i e A_{\mu}\left[Q,\Sigma\right]+i\frac{\partial_{\mu}a}{f_a}( k_L \Sigma-\Sigma  k_R) \, , 
    \end{align}
where $Q= {\rm diag} (2/3,-1/3,-1/3)$ is the electric charge matrix of the light quarks. 

Notice that the axion enters the chiral Lagrangian only through derivative terms, since there is no axion-gluon coupling. This gives the kinetic mixing with the mesons in the diagonal entries of $\Phi$, apart from the usual meson mass matrix. Defining $\phi =  (a, \pi^0,\eta_8,\eta_0)$, 
one obtains for the relevant quadratic Lagrangian ${\cal L} \supset 1/2 K_{ij} \partial_\mu \phi_i \partial^\mu \phi_j - 1/2 M^2_{ij} \phi_i \phi_j$, with 
\begin{align}
K_{ij} & = \delta_{ij} + K_{i1} \delta_{1j}  + K_{j1} \delta_{1i}  \, .
\end{align}
Here
\begin{widetext}
\begin{align}\label{Ki1}
K_{i1} & =  - \frac{\epsilon}{2 \sqrt 6} \begin{pmatrix} {\cal O} (\epsilon^2) \\ \sqrt{3} (C_u - C_d) \\ C_u + C_d - 2 C_s \\ \sqrt{2} (C_u + C_d + C_s)    
\end{pmatrix} \ , \qquad M^2=\begin{pmatrix}
m_a^2 & 0 & 0 & 0 \\
0 & 2 B_0 \hat{m}& -\frac{B_0}{\sqrt{3}}\Delta & -\sqrt{\frac{2}{3}}B_0\Delta  \\
0 &  -\frac{B_0}{\sqrt{3}}\Delta & \frac{2}{3} B_0 (\hat{m}+2 m_s) & \frac{4}{3\sqrt{2}}B_0(\hat{m}-m_s) \\
 0 &  -\sqrt{\frac{2}{3}}B_0\Delta & \frac{4}{3\sqrt{2}}B_0(\hat{m}-m_s) & \frac{2}{3} B_0 (2\hat{m}+ m_s) +M^2_0  
 \end{pmatrix} \ ,
\end{align}
\end{widetext}
where we have defined $\hat{m}\equiv (m_u+m_d)/2$, $\Delta = m_d-m_u$ and $\epsilon \equiv f_{\pi}/f_a$. 

We continue by taking the isospin limit  $\Delta = 0$, so that only $\eta_8 - \eta_0$ mass mixing takes place. The mass basis is thus defined by a single rotation 
\begin{align}
\label{eq:eta_mix}
   \begin{pmatrix}
      \eta_8\\
      \eta_0
  \end{pmatrix} 
  & =
  \begin{pmatrix}
      \cos {\theta} & \sin {\theta}\\
      - \sin {\theta} & \cos {\theta}
  \end{pmatrix} 
   \begin{pmatrix}
      \eta\\
      \eta'
  \end{pmatrix} \, , 
  \end{align}
with the rotation angle $\theta$ given by 
\begin{align}
\tan  \theta = \frac{4}{3 \sqrt 2} \frac{ B_0 (\hat m - m_s)}{m_{\eta^\prime}^2 - 2/3 B_0 (\hat{m} + 2 m_s) } \, .
\end{align}
In the limit where axion-meson mixing can be neglected, $f_a  << f_\pi$, the entries of the meson mass matrix can be obtained as usual, giving in the isospin limit $m_\pi^2 = 2 B_0 \hat{m}$ and $m_K^2 = B_0 (\hat{m} + m_s)$. The explicit $U(1)_A$ breaking term $M_0^2$ is determined by the $\eta^\prime$-mass, so that not only the mixing angle, but also the $\eta$-mass are predicted, at least in the leading-order (LO) level we are considering here. We obtain for the $\eta$-mass (in agreement with e.g. Ref.~\cite{Leutwyler:1997yr})
\begin{equation}
\label{etamass}
    m_{\eta}^2=m_{\eta_8}^2-\frac{8}{9}\frac{(m_{\pi}^2-m_K^2)^2}{m_{\eta '}^2-m_{\eta_8}^2},
\end{equation}
where  $m_{\eta_8}^2 \equiv (4 m^2_{K}-m_{\pi}^2)/3 \approx 566 \MeV$ denotes the $\eta$-mass one obtains using the Gell-Mann-Okubo formula, that is, decoupling the $\eta^\prime$. Instead Eq.~\eqref{etamass} gives $m_{\eta}\approx 494$ MeV, which is in mild tension with the measured $m_{\eta}\approx 548$ MeV. It is well known that $\chi$PT at leading order is not adequate to describe $\eta-\eta^\prime$ mixing~\cite{Georgi:1993jn, Gerard:2004gx}, and ${\cal O}(p^4)$ give important corrections to mixing angles and masses~\cite{Leutwyler:1997yr, Beisert:2001qb, Alves:2017avw}. For our purposes however  the LO result for the mixing angle suffices, keeping in mind that uncertainties from higher-order corrections are large. This gives in agreement with Ref.~\cite{Leutwyler:1997yr}
\begin{align}
\tan  \theta = \frac{4}{3 \sqrt 2} \frac{ m_\pi^2 - m_K^2}{m_{\eta^\prime}^2 -  m_{\eta_8}^2} \, ,
\end{align}
and numerically $\theta\approx -20^{\circ}$. Other methods give values ranging from $ -13^{\circ}$ to $ -22^{\circ}$~\cite{Beisert:2001qb}, so that in the following we work with  the choice $\sin \theta \approx -1/3$, as frequently done in the literature~\cite{Aloni:2018vki, Cheng:2021kjg}.
 The field redefinition in Eq.~\eqref{eq:eta_mix} modifies kinetic mixing, resulting in a rotation acting on $K_{i1}$ in Eq.~\eqref{Ki1}, which becomes after setting $s_\theta \equiv \sin \theta \approx -1/3, c_\theta \equiv \cos \theta \approx 2 \sqrt{2}/3$
\begin{align}
        K_{21} &\to -\frac{\epsilon}{2\sqrt{6}} \left[ (C_{u}+C_{d})(c_{\theta}-\sqrt{2}s_{\theta})-2 C_{s}(c_{\theta}+\frac{s_{\theta}}{\sqrt{2}}) \right] \nonumber \\
        & = - \frac{\epsilon}{2\sqrt{3}} (C_u + C_d - C_s)\nonumber \, , \\
         K_{31} &\to -\frac{\epsilon}{2\sqrt{6}} \left[ (C_{u}+C_{d})(s_{\theta}+ \sqrt{2} c_{\theta}) - 2 C_{s}(s_{\theta}-\frac{c_{\theta}}{\sqrt{2}}) \right]  \nonumber \\
         & =   - \frac{\epsilon}{2\sqrt{6}} (C_u + C_d + 2 C_s) \, .
\end{align}
Finally we canonically normalize kinetic terms, and re-diagonalize the mass matrix. At linear order in $f_a/f_\pi$ this is straighforward, and gives the following relation between the fields in the original basis $\phi =  (a, \pi^0,\eta_8,\eta_0)$ and canonically normalized mass eigenstates $\phi_{\rm phys} =  (a_{\rm phys} , \pi^0_{\rm phys} ,\eta_{\rm phys} ,\eta^\prime_{\rm phys} )$ 
\begin{align}
\label{eq:mixing}
           \pi^0 &\approx\pi^0_{\rm phys}+\epsilon \frac{C_u-C_d}{2 \sqrt{2}}\frac{m_a^2}{m_a^2-m_{\pi}^2} a_{\rm phys} \, , \\
        \eta_8 &\approx\eta_{\rm phys}+ \epsilon \frac{C_u + C_d - C_s}{2 \sqrt{3}}  \frac{m_a^2}{m_a^2-m_{\eta}^2} a_{\rm phys} \, , \\
        \eta_0 &\approx \eta'_{\rm phys}+ \epsilon \frac{C_u + C_d + 2 C_s}{2 \sqrt{6}} \frac{m_a^2}{m_a^2-m_{\eta'}^2}  a_{\rm phys} \, .
   \end{align}
These results allow  to compute the light quark contribution to the axion couplings to photons, which the axion inherits from the meson couplings, suppressed by mixing. The pseudoscalar couplings to photons induced by the electromagnetic anomaly read
\begin{align}
         \mathcal{L}_{\rm EMA}&=\frac{i}{2}\frac{N_c \alpha_{\rm em}}{4\pi} F_{\mu \nu}\widetilde{F}^{\mu \nu} {\rm Tr} \, [Q^2(\log{\Sigma}-\log{\Sigma^\dagger})] \nonumber \\
       & = -\frac{\alpha_{\rm em}}{4\pi f_{\pi}} F \widetilde{F} \left(\sqrt{2}\pi^0+\sqrt{\frac{2}{3}}\eta_8+\frac{4}{\sqrt{3}}\eta_0\right) \, .
 \end{align}
 Plugging in the mixing relations in Eq.~\eqref{eq:mixing}, and matching to the axion-photon coupling defined as
 \begin{align}
           \mathcal{L}_{\rm a \gamma \gamma}&= C_{\gamma\gamma} \frac{\alpha_{\rm em}}{4 \pi}\frac{a}{f_a} F_{\mu \nu}\widetilde{F}^{\mu \nu} \, , 
         \end{align}
 we finally obtain the light quark contribution $C^{\rm light}_{\gamma\gamma}$ in Eq.~\eqref{light}, which is  valid in the limit $m_a\ll m_{\pi,\eta,\eta'}$ and taking $s_{\theta}\approx -1/3$.

\bibliographystyle{JHEP}
\bibliography{bib}

\end{document}